\documentclass[%
 reprint,
 amsmath,amssymb,
 aps,
prb,
]{revtex4-2}

\usepackage{graphicx}
\usepackage{dcolumn}
\usepackage{bm}

\usepackage{graphicx}
\usepackage{hyperref}
\usepackage{cleveref}     
\crefformat{figure}{Fig.~\protect{#2#1#3}} 
\Crefformat{figure}{Figs.~\protect{#2#1#3}} 
\crefformat{section}{Sec.~\protect{#2#1#3}} 
\crefformat{equation}{Eq.~\protect{#2#1#3}} 

\usepackage[separate-uncertainty = true]{siunitx} 
\newcommand{\vbias}{V_{\mathrm{bias}}}
\newcommand{\vmod}{V_{\mathrm{mod}}}
\newcommand{\vstab}{V_{\mathrm{stab}}}
\newcommand{\istab}{I_{\mathrm{stab}}}

\newcommand{\didu}{$\mathrm{d}I/\mathrm{d}V$-spectrum}
\newcommand{\didus}{$\mathrm{d}I/\mathrm{d}V$-spectra}
\newcommand{\didusig}{$\mathrm{d}I/\mathrm{d}V$}

\newcommand{\deeo}{$[1 \overline{1}0]$-direction}
\newcommand{\dooe}{$[001]$-direction}

\newcommand{\eeo}{$[1 \overline{1}0]$}
\newcommand{\ooe}{$[001]$}
\newcommand{\eee}{$[1\overline{1}1]$}

\begin{document}


\title{Systematic study of Mn atoms, artificial dimers and chains\\on superconducting Ta(110)}

\author{Philip Beck}%
\affiliation{Department of Physics, University of Hamburg, Hamburg, Germany}
\author{Lucas Schneider}%
\affiliation{Department of Physics, University of Hamburg, Hamburg, Germany}

\author{Roland Wiesendanger}%
\affiliation{Department of Physics, University of Hamburg, Hamburg, Germany}
\author{Jens Wiebe}
\email{jwiebe@physnet.uni-hamburg.de}
\affiliation{Department of Physics, University of Hamburg, Hamburg, Germany}
\bibliographystyle{apsrev4-2}

\date{\today}

\begin{abstract}
Magnetic adatoms coupled to an $s$-wave superconductor give rise to local bound states, so-called Yu-Shiba-Rusinov states. Focusing on the ultimate goal of tailoring chains of such adatoms into a topologically superconducting phase, we investigate basic building blocks -- single Fe and Mn adatoms and Mn dimers on clean superconducting Ta(110) -- using scanning tunneling microscopy and spectroscopy. We perform a systematic study of the hybridizations and splittings in dimers, and their dependence on the crystallographic directions and interatomic spacings, in order to identify potentially interesting chain geometries for this novel sample type. Subsequently, we study the spin structure as well as the length dependent Shiba band structure in Mn chains of those geometries using spin-resolved scanning tunneling spectroscopy. All results are compared to the according properties of structurally identical dimers and chains on the previously studied Nb(110), which has almost identical surface structure and electronic properties, but an about three times smaller spin-orbit interaction.
\end{abstract}

\maketitle


\section{\label{sec:intro}Introduction}
The local pair-breaking potential of magnetic adatoms or embedded impurities placed on $s$-wave superconductors gives rise to local bound states, so-called Yu-Shiba-Rusinov (YSR) states \cite{Yu1968, Shiba1968, Rusinov1969, Rusinov1969a}. Following their first experimental observation \cite{Yazdani1997} using scanning tunneling microscopy (STM) and spectroscopy (STS), a plethora of physical properties have been investigated experimentally, including quantum phase transitions of YSR states by tuning the exchange coupling to the superconducting substrate \cite{Farinacci2018, Huang2020comm}, the spin polarization of the YSR state \cite{Cornils2017}, the orbital structure of YSR states \cite{Ruby2016, Choi2017} and the hybridization of YSR states in impurity dimers \cite{Ruby2018, Choi2018, Kezilebieke2018}.
\newline
The latter property is the basis for the idea to use 1D arrays of magnetic impurities to tailor topological superconductivity and Majorana bound states (MBS) \cite{Oreg2010,Choy2011,Pientka2013,Nadj-Perge2013,Klinovaja2013a,Li2014,Schecter2016}. For various experimental platforms, where artificial structures were fabricated by STM tip-induced atom manipulation, studies of the single YSR impurity and dimers have proven useful to understand the properties of chains, e.g. for Fe on Re(0001) \cite{Kim2018}, Cr on $\beta-\mathrm{Bi}_2\mathrm{Pd}$ \cite{Mier2021}, Mn on Nb(110) \cite{Schneider2021b,Beck2021,Schneider2021a}, Cr on Nb(110) \cite{Kuster2021a, Kuster2021b} and Fe on $\mathrm{NbSe}_2$ \cite{Liebhaber2021}. Therefore, one can consider the in-depth study of artificial dimers as a prerequisite for tailoring the Shiba bands in chains, which is necessary to determine potentially interesting crystallographic building directions and spacings for a given adatom species and superconducting substrate. Finally, highly tunable dimers of YSR impurities and their detailed study may enable the implementation of novel qubit designs \cite{Mishra2021}.
\newline
Here, we perform a detailed study of Fe and Mn adatoms on clean Ta(110). To begin with, we introduce a novel procedure for the preparation of clean Ta(110), which captivates with an easy technical realization and less stress on UHV components, compared to previously reported methods \cite{Eelbo2016}. Additionally, we demonstrate highly reliable and reproducible STM tip-induced atom manipulation with Mn adatoms, which is a key requirement for the investigation of artificial nanostructures. 
\newline
By studying the electronic in-gap states and the spatial extent of YSR states of Mn and Fe adatoms (\cref{sec:atoms}), as well as artificially constructed Mn dimers with various spacings (\cref{sec:001dimers} and \cref{sec:110dimers}), we proceed step-by-step to identify the potentially interesting spacings for the $[001]$- and \deeo s. We come to the conclusion that close-packed Mn chains in both crystallographic directions should be interesting in terms of their in-gap electronic properties, which are exactly the geometries that have been intensively studied for the considerably less spin-orbit coupled Mn/Nb(110) system \cite{Schneider2021a, Schneider2021b, Schneider2021}. We perform a comparative study of both chain systems here (\cref{sec:001chain}, \cref{sec:110}) and in Ref.\cite{Beck2022_2}.
Last, we perform spin-polarized (SP)-STM with YSR-functionalized superconducting tips (\cref{sec:spstm110}) to measure the spin structures of both chains (that of the \dooe~can be found in Ref.\cite{Beck2022_2}), which show indications for a dominating ferromagnetic order. 
\section{\label{sec:Methods}Methods}
\subsection{\label{subsec:setup} Experimental setup and preparation of clean Ta(110)}
The measurements shown in this publication and its accompanying manuscript \cite{Beck2022_2} were performed in a home-built ultrahigh vacuum (UHV) STM system \cite{Wiebe2004}, consisting of three chambers: One designated preparation chamber, containing a home-built e-beam heater for high temperature flashes, another chamber containing a system for low-energy electron diffraction (LEED) measurements and a chamber connected to an UHV wet cryostat including a $^3$He refrigerator, which hosts an STM operated at a temperature of \SI{320}{\milli \kelvin} in a magnetic field up to \SI{12}{\tesla} perpendicular to the sample surface. We used a mechanically sharpened superconducting Nb tip for all measurements to obtain an improved energy resolution by overcoming the Fermi-Dirac distribution limit at finite temperatures, which regular metal tips would usually impose on the experiment \cite{Ji2008}. 
\newline
In contrast to previously reported methods for cleaning Ta(110) which include high temperature annealing cycles at \SI{2500}{\kelvin} \cite{Eelbo2016} for multiple hours or annealing in oxygen atmosphere \cite{Engelkamp2015}, we demonstrate that a preparation of a similar surface quality is achievable by consecutive \SI{30}{\second} long flashes at higher temperatures. The flashes were performed using an e-beam heater with the sample positioned such that the polished side of the Ta(110) single crystal (mounted on a tungsten plate using tungsten wires) faced the filament. After flashing the sample at a given power, we characterized its cleanliness by LEED and STM.
\newline
 \begin{figure}
    \centering
    \includegraphics[scale=1]{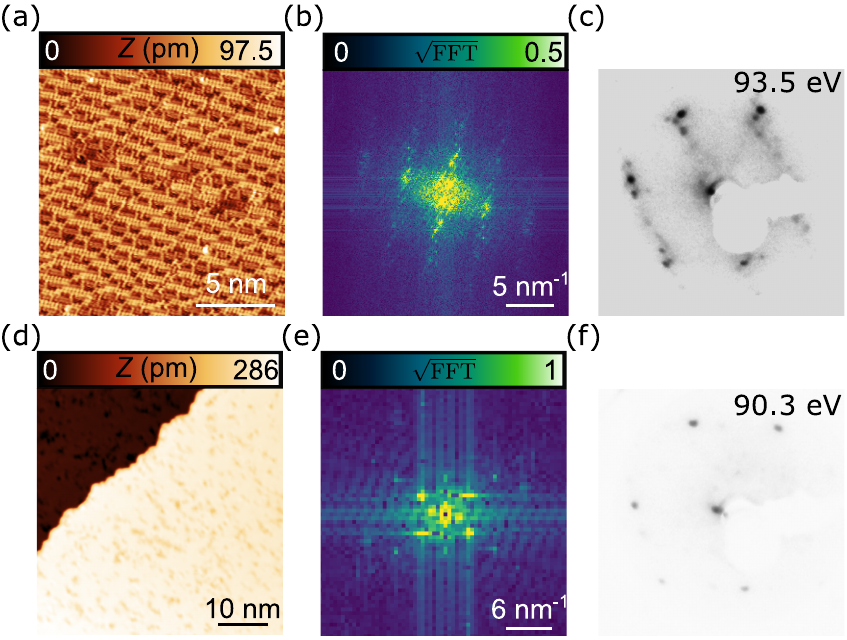}
    \caption{\label{fig:preparation} (a-c) Oxygen-reconstructed Ta(110) obtained by flashing the sample with a power of \SI{250}{\watt} characterized by (a) an overview STM image, (b) the 2D-FFT of an atomically resolved STM image and (c) the LEED pattern obtained at \SI{93.5}{\electronvolt}. (d-f) Clean Ta(110) prepared by flashing the sample with a power of \SI{360}{\watt} characterized by (d) an overview STM image, (e) the 2D-FFT of an atomically resolved STM image and (f) the LEED pattern measured with a beam energy of \SI{90.3}{\electronvolt}.  Measurement parameters: (a) and (b) $\vbias=\SI{-20}{\milli \volt}$, $I=\SI{2}{\nano \ampere}$, (d) $\vbias=\SI{5}{\milli \volt}$, $I=\SI{1}{\nano \ampere}$, (e) $\vbias=\SI{10}{\milli \volt}$, $I=\SI{1}{\nano \ampere}$.}
\end{figure}
An STM image, a 2D-FFT of an atomic resolution STM image and the LEED pattern of an exemplary sample which was flashed at a power of \SI{250}{\watt} are shown in \Cref{fig:preparation}(a)-(c). The atomically resolved overview STM image and the 2D-FFT show an oxygen-induced reconstruction of the Ta(110) surface \cite{Eelbo2016}, which looks very similar to the well-known oxygen-induced reconstruction of Nb(110) \cite{Suergers2001}. The similarity of the LEED pattern measured on this sample (\cref{fig:preparation}(c)) in comparison with LEED measurements of reconstructed Nb(110) \cite{Arfaoui2002} suggest a similar structure of the oxygen-induced reconstruction. It is composed of two distinct crystallographic layers: the Nb(110) substrate and a slightly distorted fcc layer of NbO \cite{Kuznetsov2011}.
\newline
After gradually increasing the flashing power to \SI{360}{\watt}, we observe a first change in the LEED pattern and STM images as shown in \Cref{fig:preparation}(d)-(f). The overview STM image in \cref{fig:preparation}(d) shows largely flat terraces which are covered by some defects appearing as dark spots. We interpret those as residual oxygen atoms. The impurities cover $\SI{15}{\percent}$ of the clean Ta(110) regions. The 2D-FFT of an atomic resolution STM image is shown in \cref{fig:preparation}(e). It confirms that the flat areas observed in (d) are clean Ta(110) with an unreconstructed (1x1) bcc(110) surface. Further evidence for the large-scale cleanliness of the surface is apparent from the LEED pattern shown in \cref{fig:preparation}(f), which only displays sharp diffraction spots of the bcc(110) surface in contrast to the LEED pattern of reconstructed Ta(110) in \cref{fig:preparation}(c).
\newline
By further increasing the flashing power to \SI{380}{\watt} we obtain a clean Ta(110) surface which is merely covered by $6\%$ of oxygen impurities (overview STM image in Ref.\cite{Beck2022_2}). This sample was used for all subsequent experiments with Mn and Fe adatoms and their artificial structures. It is crucial to have an extremely low defect density in order to obtain reproducible \didus, since the local environment of magnetic adatoms strongly influences the YSR states as recently shown in Ref.\cite{Odobesko2020}. It should be noted that the exact flashing power, which is required to clean Ta(110), can vary between different experimental set-ups and samples.
\subsection{\label{subsec:manipulation}STM tip-induced atom manipulation on Ta(110)}
\begin{figure}
    \centering
    \includegraphics[scale=1]{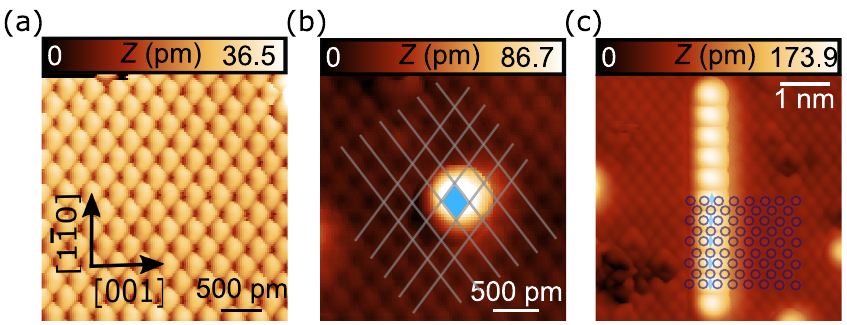}
    \caption{\label{fig:manipulation}(a) STM image obtained while stabilizing at manipulation parameters and after approaching a Mn atom, which is then bound by the tip-potential and used to scan the surface of clean Ta(110). Black arrows and labels indicate crystallographic directions, which are valid for all three panels. (b) Atomically resolved STM images displaying a single Mn atom and (c) an artificially constructed close-packed Mn chain in \deeo. White lines in (b) highlight the $[1\overline{1}1]$-oriented atomic rows of Ta atoms and reveal the adsorption geometry of the Mn atom (blue diamond). Dark blue circles in (c) highlight the atoms of the Ta(110) surface and blue diamonds mark the fourfold-coordinated adsorption sites on which the Mn adatoms of the chain are positioned. Measurement parameters: (a) $\vbias=\SI{-2.2}{\milli \volt}$, $I=\SI{121}{\nano \ampere}$, (b) and (c) $\vbias=\SI{10}{\milli \volt}$, $I=\SI{1}{\nano \ampere}$.}
    
\end{figure}
We deposited single Mn atoms and Fe atoms onto this Ta(110) surface in two steps: We started with the deposition of Mn atoms and subsequently carried out all investigations of single Mn atoms and artificial structures. Afterwards we additionally evaporated Fe atoms onto the sample, which allows us to clearly distinguish both adatom species and their properties. Both depositions were carried out while maintaining a sample temperature $< \SI{6}{\kelvin}$ to achieve statistically distributed single adatoms (overview image in Ref.\cite{Beck2022_2} and \cref{fig:atomspec}(a), see below).
\newline 
Mn atoms were reliably positioned by lateral STM tip-induced atom manipulation \cite{Eigler1990} at typical tunneling resistances of $\sim \SI{30}{\kilo \ohm}$, depending on the specific microtip. Approaching a single Mn atom with the STM tip stabilized at manipulation parameters and subsequently scanning a clean area of Ta(110) while manipulating the Mn atom yields a manipulation image \cite{Stroscio2004} as shown in \cref{fig:manipulation}(a). The obtained image of equally favored adsorption sites matches an atomic resolution image of the clean Ta(110) surface regarding the symmetry and spacing between equal sites. We conclude that there is only a single stable adsorption site for Mn adatoms on Ta(110). An atomically resolved STM image including a Mn atom is shown in \cref{fig:manipulation}(b). White lines highlight the $[1\overline{1}1]$-directions of atomic rows. By evaluating the location of the Mn adatom in the resulting grid of the white lines, we conclude that Mn atoms are adsorbed in the fourfold-coordinated hollow site of Ta(110).   
\newline
Using the atom manipulation image shown in \cref{fig:manipulation}(a) as an overlay, we are able to construct nanostructures composed of atomically precise positioned Mn atoms. An example of such an artificial structure is shown in \cref{fig:manipulation}(c), which displays a Mn chain consisting of eleven atoms oriented in the \deeo~with the close-packed spacing of $\sqrt{2} a$, where $a$ is the lattice constant of bcc Ta (\SI{331}{\pico \meter}) \cite{Ashcroft2012}. Blue circles are an overlay of the atomic resolution of the Ta(110) surface which one can clearly see, e.g. in the top left corner. Blue diamonds mark the positions of the fourfold-coordinated hollow sites, in which one would expect the Mn atoms of the chain to be positioned, based on a continuation of the Ta lattice. We find very good agreement of the Mn atoms' positions with the blue diamonds, indicating that the fabricated structure is indeed a close-packed chain in \deeo~with homogeneous spacing. This is an important finding concerning the perfect geometry of our artificially constructed chains where all Mn atoms in the bulk of the chain are adsorbed/coordinated equally and, therefore, experience similar interactions with the physical environment e.g. coupling strength to Ta(110) substrate electrons or overlap of the YSR states with neighboring Mn atoms.
\newline
While it is not necessarily possible to manipulate dimers or chains in all crystallographic directions or distances for all adatom-substrate-combinations, we find that this is reliably possible for Mn on Ta(110) in all three symmetric directions \ooe, \eeo~ and \eee~ in every spacing down to the close-packed distances $a$, $\sqrt{2}a$ and $(\sqrt{3}/2) a$, respectively. On the other hand, we found that it is tremendously difficult to manipulate Fe atoms while maintaining a stable tip, which rendered it impossible to create artificial structures.
\subsection{\label{sec:protocol} STM and STS measurements}
We refer to STM images as scans obtained by the following procedure, commonly known as constant-current images: We applied a bias voltage $V_{\mathrm{bias}}$ to the sample upon which the tip-sample distance is controlled by a feedback-loop such that a constant current $I$ is achieved.
\newline
\didus~were obtained by standard lock-in technique using a modulation frequency of $f_{\mathrm{mod}}=\SI{4142}{\hertz}$ and a modulation amplitude referred to as $\vmod$ with a typical value of $\SI{20}{\micro \volt}$ (rms value) added to $\vbias$. Prior to a \didu, the tip was stabilized at $\vstab$ and $\istab$. After an initial settling time, the feedback-loop was turned off and the bias was swept through a defined range.
\newline
\didusig-grids were obtained by recording \didus~on a predefined spatial grid, which was positioned over the structure of interest. \didusig -maps are the 2D maps of a slice of the grid evaluated at a given bias voltage. Typical measurement parameters are of the same size as for individual \didus.
\didusig-line profiles are measured similarly to \didusig-grids, with the exception that the spatial grid which is positioned over the structure of interest is one-dimensional. Constant-contour \didusig-maps were obtained by repeated scanning of individual lines of STM images. In a first sweep, each line is measured as it would be the case in a regular STM image. The $z$-signal of this sweep is saved. In the next sweep, the bias voltage $\vbias$ is set to a specific value, for which one aims to obtain a \didusig-map. The previously recorded $z$-signal is retraced, while the actual feedback is turned off. This allows the measurement of \didusig -maps at biases located in the superconducting gap of the sample, which would not be possible using conventional STM images.
\newline 
Due to the use of a superconducting tip, spectroscopic features of the sample are shifted by $\pm \varDelta _ {\mathrm{tip}}$ in the raw data of \didus. To display all features at their actual energy, we process the data by numerical deconvolution as described in Ref.\cite{Pillet2010}. As an input for this method we describe the tip's density of states using a Dynes function \cite{Dynes1978}, which in return requires knowledge about the superconducting gap of the tip. By analyzing multiple Andreev reflections \cite{Ternes2006} in \didus~ obtained with a small tip-sample distance, which we acquired prior to most measurements shown here, we can accurately determine this value. Further input parameters are the broadening $\gamma = \SI{1e-6}{\electronvolt}$ and the measurement temperature $T=\SI{320}{\milli \kelvin}$.
\newline
Spin-polarized STM tips, which were used for the SP-STM and -STS measurements, were prepared by picking up individual Mn adatoms with the superconducting Nb tip, giving rise to YSR states on the tip apex. To verify the magnetic sensitivity of a correspondingly prepared tip, we performed STM measurements over an AFM coupled $(\sqrt{3}/2) a-[1\overline{1}1]$ Mn chain, where alternating bright-dark contrast was easily achievable~\cite{Schneider2021}.
\newline
To determine the size of superconducting gaps, such as the minigap observed in Ref.\cite{Beck2022_2}, we fit the relevant energy range of a given \didus~using a Dynes function\cite{Dynes1978}
\begin{equation}
    \label{eq:Dynes}
    N_{\mathrm{SC}}(E) = N_{N}(E_{\mathrm{F}}) \mathrm{Re}(\frac{|E| + i\gamma}{\sqrt{(|E|+i \gamma)^2-\varDelta^2}}),
\end{equation}
where $N_{N}(E_{\mathrm{F}})$ is the normal state density of states at the Fermi level.
\section{\label{sec:results}Results}
We commence this chapter with the investigation of single Mn and Fe atoms on clean Ta(110) in \cref{sec:atoms}. Following this, we move on to study artificial dimers in \cref{sec:001dimers} and \cref{sec:110dimers}, where we restrict ourselves to structures composed of Mn atoms, as the spectroscopic properties and the incapability to manipulate Fe atoms make this species inappropriate for 1D systems anyhow. Having learned about the distance- and direction-dependent interaction of Mn atoms we move on to investigate the potentially interesting Mn chains in regards of their Shiba bands, where we focus on spectroscopic data on chains in \dooe~with an interatomic spacing of $2a$ (\cref{sec:001chain}) and chains in \deeo~with a spacing of $\sqrt{2}a$ (\cref{sec:110}). We proceed to perform SP-STS employing a YSR functionalized tip to study the potentially complicated magnetic ground state of chains in \deeo~with $\sqrt{2}a$ spacing (\cref{sec:spstm110}).
\subsection{\label{sec:atoms}Spectroscopic investigation of Mn and Fe atoms}
\begin{figure*}
    \centering
    \includegraphics[scale=1]{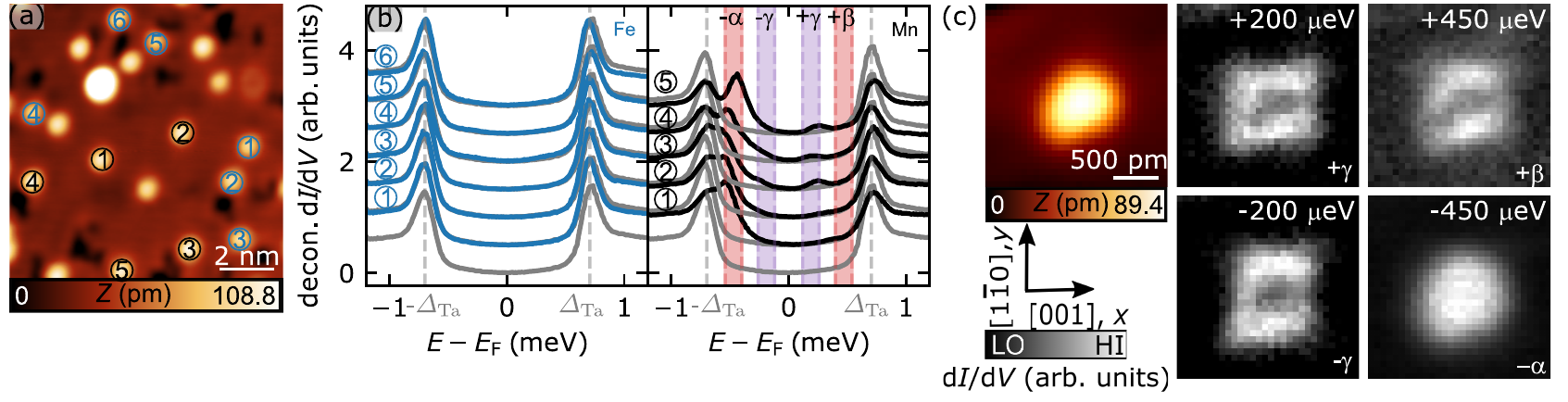}
    \caption{\label{fig:atomspec}(a) STM overview image showing statistically distributed Mn and Fe atoms. Color-coded encircled numbers highlight the positions where \didus~were measured. (b) \didus~measured on Fe atoms (blue), on Mn atoms (black) and on the Ta(110) substrate (gray). Color-coded encircled numbers link the spectra to a position in the overview image shown in (a). A substrate \didu~is pasted behind every other spectrum to enable an easy comparison. Spectra are offset by 0.5~arb.units for the sake of visibility. Gray vertical lines mark the coherence peaks of Ta. Red- and purple-colored lines enclose shaded energy areas where YSR states of randomly positioned Mn atoms are typically found. (c) Spectroscopic grid of a single isolated Mn impurity shown in the top left panel. Numbers in the top right corner indicate the energy slice. Crystallographic directions valid for all panels of (c) are given in the bottom left corner. Measurement parameters: (a) $\vbias=\SI{-20}{\milli \volt}$, $I=\SI{200}{\pico \ampere}$, (b) and (c) $\vstab=\SI{-2.5}{\milli \volt}$, $\istab=\SI{1}{\nano \ampere}$ and $\vmod=\SI{20}{\micro \volt}$.}
\end{figure*}
An overview STM image of a Ta(110) sample covered with randomly distributed Fe and Mn atoms is shown in \cref{fig:atomspec}(a). A closer look at the adatoms, which are visible as bright protrusions, reveals that there are two dominant species present: At the measurement parameters of \cref{fig:atomspec}(a), the slightly shallower ones are identified as Fe atoms ($\SI{58}{\pico \meter}$), whereas the ones with a slightly larger apparent height ($\SI{74.3}{\pico \meter}$) are identified as Mn atoms (known from evaporation order discussed in \cref{subsec:manipulation}). A selection of each adatom species is marked by blue and black encircled numbers, respectively. Deconvoluted \didus~measured directly on top of each of the highlighted impurities are shown alongside the same color-coded numbers in \cref{fig:atomspec}(b).
A deconvoluted \didu~measured on bare Ta(110) is pasted behind every spectrum of the impurities and is displayed in gray at the very bottom of both panels of \cref{fig:atomspec}(b). We observe sharp coherence peaks located at $\varDelta_{\mathrm{Ta}} = \pm \SI{690}{\micro \electronvolt}$, which are marked by gray dashed vertical lines. At our measurement temperature of $\SI{320}{\milli \kelvin}$ the superconducting gap of bare Ta(110) is fully pronounced, i.e. the \didusig-signal approaches the noise level at energies lower than $\varDelta_{\mathrm{Ta}}$, as can be seen in the data in Ref.\cite{Beck2022_2}. $\varDelta_{\mathrm{Ta}}$ matches reference values \cite{Townsend1962}. However, for some of the used tips, a residual \didusig-signal appears inside the gap, e.g., in \cref{fig:atomspec}(b). This is most probably the result of a tip with a slightly reduced gap value or quasiparticle lifetime.
\newline
A comparison of the six \didus~taken on Fe atoms with the gray-colored substrate spectra pasted behind each of the curves reveals very little differences. Most importantly, we do not observe any YSR states located between the coherence peaks of the Ta substrate. On the other hand, moving to Mn impurities, we find that the \didus~vary strongly in comparison to the substrate spectra. The coherence peaks of tantalum are reduced in intensity and we observe two pairs of YSR states inside of the gap. All spectra measured on a Mn impurity show similar features: The most intense peaks are found at $\SI{-470(70)}{\micro \electronvolt}$, as indicated by the red shaded region termed $-\alpha$. The corresponding states at positive bias termed $+\beta$ (as we will argue in the following, the states have distinct orbital origin) are very low in intensity and are not visible in the spectra shown in \cref{fig:atomspec}(b). Considering the 2D maps in \cref{fig:atomspec}(c), we will proceed to show that this is partly caused by the spatial distribution of this particular state, which has a nodal line in the center of the atom. Nevertheless, even if one averages a spectrum over the adatom extent, the YSR state at $+\SI{450(70)}{\micro \electronvolt}$ only shows a low \didusig-signal, which is most probably caused by a non-magnetic scattering contribution of the impurity \cite{Balatsky2006, Franke2011}. The second pair of YSR states is located at $\pm \SI{190(70)}{\micro \electronvolt}$ as indicated by the energy range shaded in purple and termed $\pm \gamma$.
\newline
In the following, we discuss the absence and presence of YSR states inside the superconducting gap of Ta(110) for Fe and Mn adatoms, respectively, in light of recent experiments and theoretical considerations. In a model for quasi-classical spins coupled to an $s$-wave superconducting host, which turned out to be a decent approximation for transition metal atoms having magnetic moments considerably larger than $1 \mu_\textrm{B}$, the energy of YSR states is given by 
\begin{equation}
    \label{eq:ysr_energy}
    \epsilon = \varDelta \frac{1-(J S_{\mathrm{imp}} \pi \rho_s)^2}{1+(J S_{\mathrm{imp}} \pi \rho_s)^2}, 
\end{equation}
where $J$ is the exchange coupling strength of the local impurity spin of size $S_{\mathrm{imp}}$ to the substrate conduction electrons and $\rho_s$ is the substrate electron density of states evaluated at the Fermi level in the normal state \cite{Yu1968, Shiba1968, Rusinov1969, Rusinov1969a}. Recent experiments and theoretical calculations have shown that $J$ is subject to a systematic trend regarding the $3d$ transition metal series \cite{Schneider2019, Kuster2021}. Starting with the atomic species of Mn, $J$ is expected to increase for higher and lower atomic number species. Exactly this trend has been shown to hold true for several $3d$ transition metal species on clean Nb(110) \cite{Odobesko2020, Beck2021, Kuster2021}. We, therefore, speculate, that $J S_{\mathrm{imp}} \pi \rho_s$ is considerably increased for Fe with respect to Mn and, according to \cref{eq:ysr_energy}, that its YSR states may be shifted into the coherence peaks of the substrate.
\newline
Considering the YSR states of Mn adatoms, one finds that the YSR states $+\alpha$ and $\pm \gamma$ vary slightly in energy for the different impurities investigated in \Cref{fig:atomspec}(a) and (b). This is naturally explained by the random distribution of the Mn atoms investigated here, which were not moved to a defect free region of the surface prior to the measurement. It has been shown, that one may influence the parameter $J S_{\mathrm{imp}} \pi \rho_s$ in \cref{eq:ysr_energy} experimentally i.e. by changing the tip-impurity distance and, thereby, modifying attractive/repulsive forces between tip and impurity \cite{Farinacci2018, Huang2020comm}, by having magnetic molecules forming a self-organized island with Moiré-like variations \cite{Hatter2015} or by bringing adatoms close to residual impurities \cite{Odobesko2020}. The latter case could be what we are observing here, since some of the Mn adatoms are quite obviously positioned in proximity to residual oxygen (3--5) or might even be located on top of a patch of it (1--2). However, it is important to note that even though the energies of the YSR states vary slightly for randomly distributed Mn adatoms, the overall spectroscopic features appear to be quite similar, i.e. two pairs of YSR states are found. Once an adatom is moved to a defect-free region, we find very reproducible YSR state energies of $\pm \SI{450}{\micro \electronvolt}$ and $\pm \SI{200}{\micro \electronvolt}$, which is crucial for the purpose of studying dimers or chains. Therefore, defect-free regions were chosen to perform the following experiments as well as the measurements in Ref.\cite{Beck2022_2} to achieve a high reproducibility.
\newline
An example of such a Mn atom in a defect free environment with YSR states of $+\alpha =+\SI{450}{\micro \electronvolt}$, $-\beta =-\SI{450}{\micro \electronvolt}$ and $\pm \gamma = \pm \SI{200}{\micro \electronvolt}$ is shown in \cref{fig:atomspec}(c). The leftmost panel is an STM image of the adatom, while the other panels are \didusig-maps \cite{Ruby2016, Choi2017}. 
As governed by the $C_{2v}$ point group of an adatom adsorbed in a fourfold-coordinated hollow site of a bcc(110) surface \cite{Beck2021}, the $d_{xy}$-, $d_{xz}$- and $d_{yz}$-orbitals belong to different irreducible representations of the group and should, therefore, have one scattering channel each. On the other hand, the $d_{x^2-y^2}$- and $d_{z^2}$-orbitals may hybridize as they belong to the same irreducible representation.
\newline
Indeed, we find spatial distributions which match the 2D projection of atomic $d$-orbitals as projected onto the $x-y$-plane. 
However, we find that the particle-hole partners of the states at $\pm \SI{450}{\micro \electronvolt}$ have distinct spatial distributions. While the one at negative biases resembles the $d_{z^2}$-orbital with slight contributions from $d_{x^2-y^2}$, the positive bias state matches a $d_{yz}$-orbital. As these states may not hybridize in an individual atom due to symmetry constraints, we interpret them as two YSR states which are closer in energy than our experimental resolution allows us to distinguish. Along with a strong intensity asymmetry, which is given by a non-magnetic scattering term, the different spatial distributions for positive and negative biases explain the intensity asymmetries of these YSR states in \cref{fig:atomspec}(b). On the other hand, the spatial distribution of the $\pm \gamma$-states are similar for positive and negative biases. They resemble the 2D projection of a $d_{xy}$-orbital, possibly with contributions from $d_{xz}$ and $d_{yz}$. Interestingly, the order of the YSR states of $d_{xy}$, $d_{xz}$, and $d_{yz}$ symmetries is changed with respect to the case of the Nb(110) substrate.
\subsection{\label{sec:001dimers}Spectroscopic investigation of Mn dimers in \dooe}
\begin{figure*}
    \centering
    \includegraphics[scale=1]{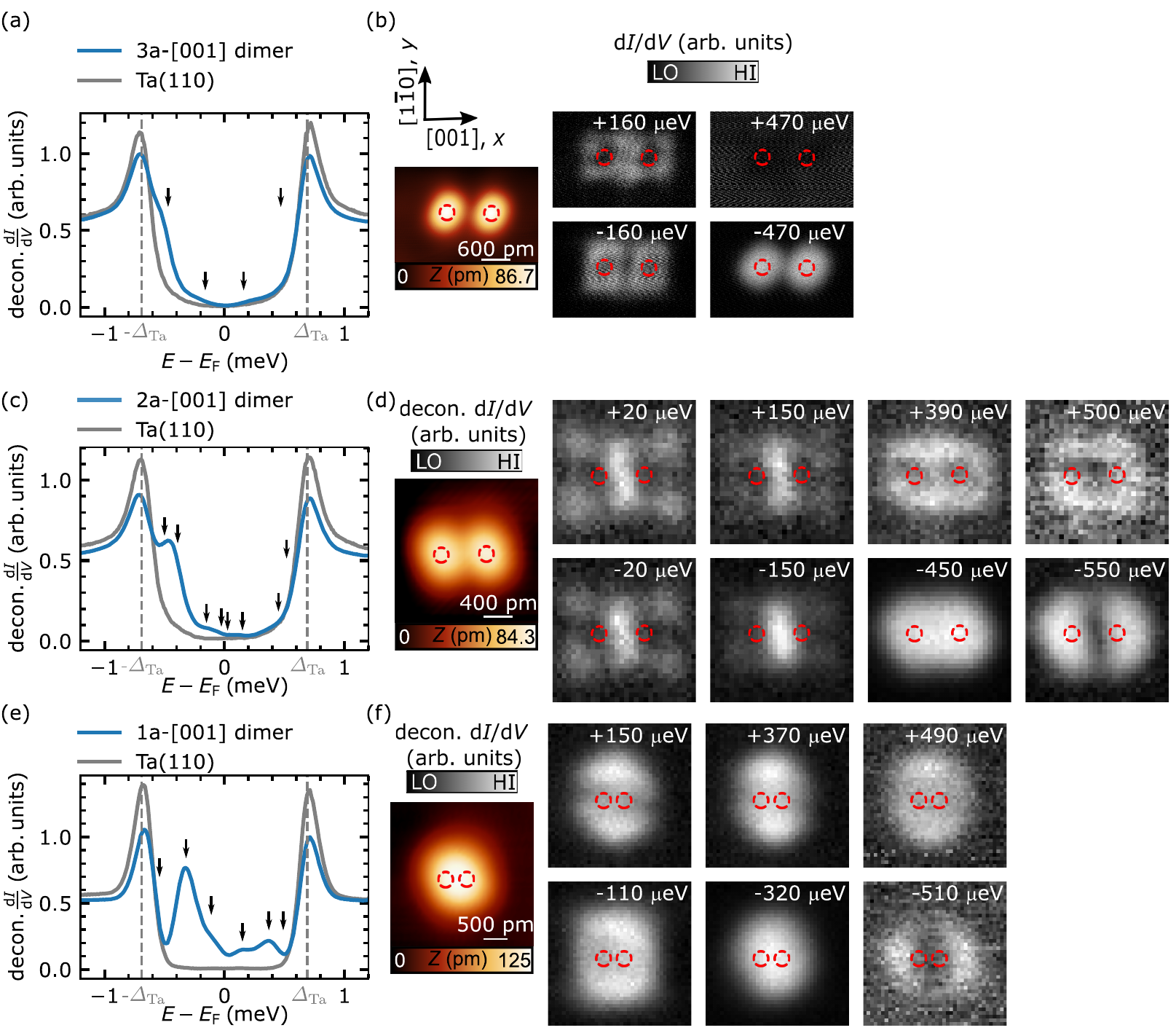}
    \caption{\label{fig:001dimers} Interatomic spacing-dependent investigation of artificial Mn dimers in \dooe~with separations of (a),(b) $3a$, (c), (d) $2a$ and (e), (f) $a$. Panels (a), (c) and (e) each show a deconvoluted \didu~of the Ta(110) substrate (gray) and a deconvoluted \didu~averaged over the spatial extent of the respective dimer (blue), in order to capture all YSR states (gray scale images, whose energies are marked by black arrows). Gray dashed vertical lines mark the superconducting gap of Ta(110). Panels (b), (d) and (f) display the spatial distributions of the YSR states, identified in the \didus. While (b) is a constant-contour \didusig-map, the measurements shown in (d) and (f) are spectroscopic grids (details in \cref{sec:protocol}). An STM image of the respective dimer is shown in the left of each panel (colored images). The crystallographic directions indicated in (b) are valid for all panels. Red circles highlight the positions of the Mn adatoms in the dimers. Measurement parameters: (a), (c), (d), (e) and (f) $\vstab=\SI{-2.5}{\milli \volt}$, $\istab=\SI{1}{\nano \ampere}$ and $\vmod= \SI{20}{\micro \volt}$, (b) $\vstab=\SI{-2.5}{\milli \volt}$, $\istab=\SI{1}{\nano \ampere}$ and $\vmod= \SI{40}{\micro \volt}$.}
\end{figure*}
Motivated by the observation of low-lying YSR states only for single Mn adatoms, we identify artificial structures constructed from Mn atoms to be more promising than those of Fe in regards of tailoring MBS. Considering the band formation in chains of hybridizing YSR atoms, it is favorable to start with single atom YSR states close to the Fermi level, as they require a smaller bandwidth to cross the Fermi level, as opposed to YSR states at $\pm \varDelta_{\mathrm{Ta}}$ which require a very large bandwidth. In order to determine potentially interesting building directions and interatomic spacings, one can study various dimers and roughly extrapolate the experimental findings to estimate the expected YSR band structure in 1D chains \cite{Ruby2018, Choi2018, Ding2021, Friedrich2021, Choi2021, Kuster2021a}. Moreover, these results can then be compared to the case of Nb(110) for which similarly oriented and separated Mn dimers have been studied \cite{Beck2021}.
\newline
A spectroscopic investigation of Mn dimers in \dooe~is shown in \cref{fig:001dimers}. Deconvoluted \didus, averaged over the spatial extent of the dimers with interatomic spacings of $3a$, $2a$ and $a$, are shown alongside a deconvoluted \didu~of the Ta(110) substrate in \Cref{fig:001dimers}(a),(c) and (e), respectively. 2D maps of the YSR states of the respective dimers are shown in \Cref{fig:001dimers}(b), (d) and (f).
\newline
A comparison of the \didu~averaged over a Mn $3a-[001]$-dimer with the spectrum of a single Mn atom free from perturbations (\cref{fig:atomspec}(b) and \cite{Beck2022_2}) shows very little differences. We still observe two pairs of peaks in the in-gap spectrum, caused by YSR states: One at $\pm \SI{470}{\micro \electronvolt}$ and one at $\pm \SI{220}{\micro \electronvolt}$, which we refer to as $-\alpha$ and $+\beta$ and $\pm \gamma$, respectively, as in the single atom case. Apart form a slight shift in energy and a different intensity distribution (due to the averaging of the spectra on the dimer in \cref{fig:001dimers}(a)), the spectroscopic properties are unchanged, indicating that the overlap of the YSR states is too low to enable a measurable hybridization and splitting. This finding is further supported by the spatial distributions of the YSR states shown in \cref{fig:001dimers}(b), which match those of two non-interacting single impurities, c.f. \cref{fig:atomspec}(c).
\newline
The situation is altered if the Mn atoms are brought closer together by one site in \dooe, resulting in a $2a-[001]$-dimer, shown in \Cref{fig:001dimers}(c) and (d). From a comparison of the \didu~of the $2a-[001]$ dimer with the $3a-[001]$ dimer, we find that the YSR states have shifted to lower energies and that they appear broader. From the \didusig-grids in \cref{fig:001dimers}(d), we find a distinguished spatial shape in the spectroscopic grids at $+\SI{390}{\micro \electronvolt}$ and $-\SI{450}{\micro \electronvolt}$ which have an increased intensity between the two Mn atoms, and different shapes at $+\SI{500}{\micro \electronvolt}$ and $-\SI{550}{\micro \electronvolt}$, with nodal lines between the two impurities. We interpret those states as symmetric and antisymmetric combinations of the single atom $\alpha$ and $\beta$ states \cite{Flatte2000, Morr2003, Morr2006}. Note that the spatial distributions of the states at negative (positive) bias resemble those of linear combinations of the single adatom $d_{z^2}$-like $\alpha$ state ($d_{yz}$-like $\beta$ state), c.f. \cref{fig:atomspec}(c). It appears that the $\alpha$ and the $\beta$ states are no longer degenerate within our energy resolution, as visible from the energies of the maps. Furthermore, we find indications that the single atom $\gamma$-state could as well be split into two YSR states: One state is observed at $\pm \SI{150}{\micro \electronvolt}$, which has an intensity maximum between the two Mn atoms and another state is found at $\pm \SI{20}{\micro \electronvolt}$ which has a similar spatial distribution between the atoms, but with prominent lobes in the \deeo~slightly shifted outwards from each of the impurities. We speculate that the high-intensity state between both atoms ($\pm \SI{150}{\micro \electronvolt}$) is still visible in the 2D map of the low-lying YSR state, due to the limited energy resolution. We, therefore, interpret those states as symmetric and antisymmetric combinations of the single atoms' $\gamma$-states. 
\newline
Judging from these results, a chain in \dooe~with a spacing of $2a$ is potentially interesting in terms of tailoring MBS, as one would start the band formation with a pair of energetically low-lying YSR states, which would only require a low bandwidth to cross the Fermi level. 
\newline 
Moving on to the close-packed $1a$ Mn dimer in \dooe, we find an altered spectrum and spatial distribution again, as illustrated in \Cref{fig:001dimers}(e) and (f). The $-\alpha$ state, which is characterized by the highest intensity in \didus, has shifted to significantly lower energies. A state with an intensity maximum in the center of the Mn atoms is found at $-\SI{320}{\micro \electronvolt}$. Additionally, we observe a state with a nodal line in the center of the dimer at $-\SI{510}{\micro \electronvolt}$. We explain those states as hybridized and split $\alpha$ states. The two spatial shapes at $+\SI{490}{\micro \electronvolt}$ and at $+\SI{370}{\micro \electronvolt}$ match the single atom $\beta$ state, as they have a dumbbell-like spatial distribution oriented in the \deeo, indicating that it is hybridized and split as well.
\newline
Unfortunately, neither the \didu~nor the maps of the YSR states give a clear picture of the energetically low-lying $\pm \gamma$-state. We only observe a pair of states at \SI{+150}{\micro \electronvolt} and \SI{-110}{\micro \electronvolt}, which resemble the spatial distribution of the single atom $\gamma$ state in the sense that they have lobes along the \deeo~and an intensity minimum at the connecting axis of the dimer. We assume that our energy resolution is too low, in order to accurately determine the energy split and shift of the single atom $\gamma$ state. A detailed discussion of the Shiba band properties and SP-STM results of a close-packed Mn chain in \dooe~are given in Ref.\cite{Beck2022_2}.
\subsection{\label{sec:001chain}Spectroscopic investigation of $2a$-\ooe~Mn chains}
\begin{figure}
    \centering
    \includegraphics[scale=1]{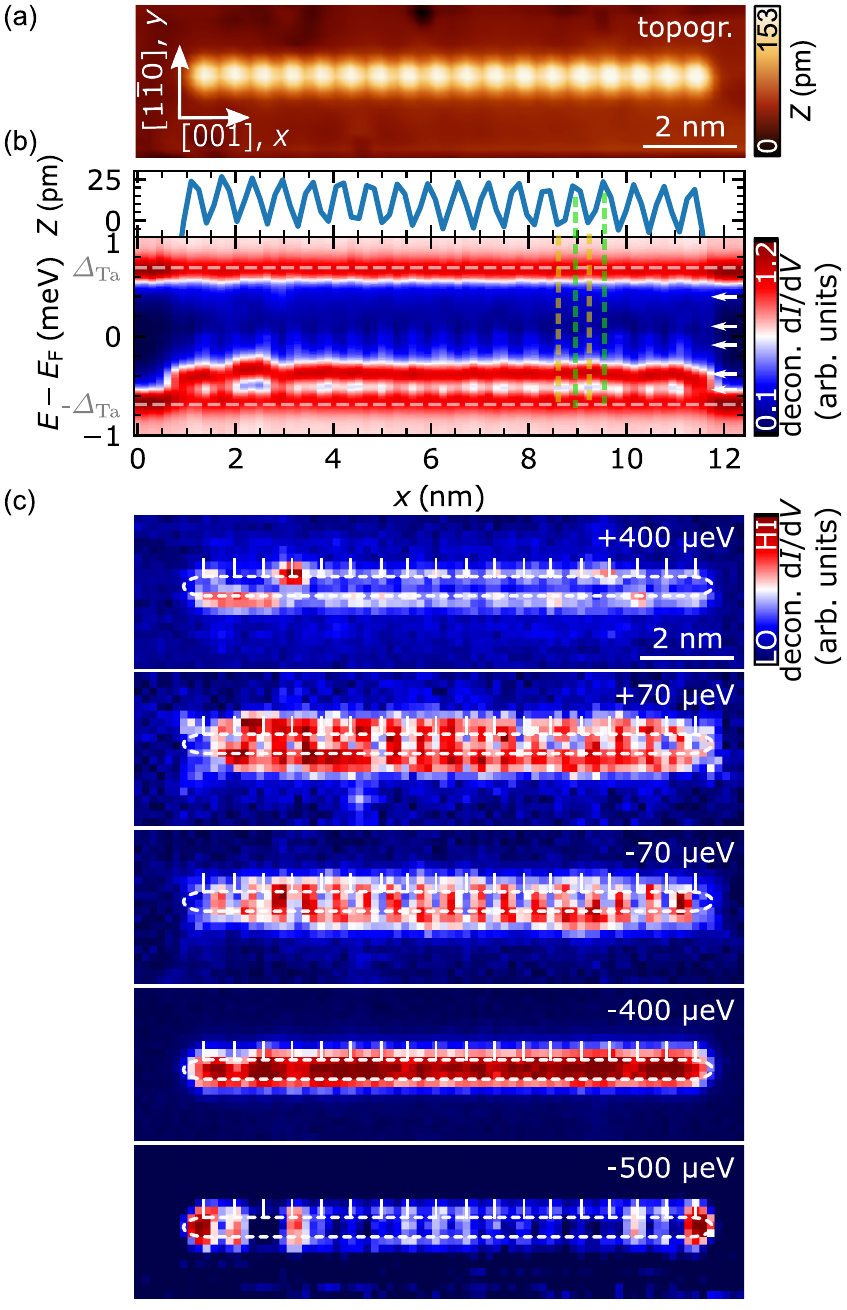}
    \caption{\label{fig:2a-001}(a) STM image of an 18 atoms long Mn chain in \dooe~with an interatomic spacing of $2a$. White arrows and labels indicate crystallographic directions, which are valid for all panels. (b) Height profile (top) and \didusig-line profile (bottom) measured along the longitudinal axis of the Mn chain shown in (a). White dashed horizontal lines mark the coherence peaks of Ta(110). Green and yellow dashed vertical lines mark examples of the locations of atomic and interatomic sites in the \didusig-line profile, respectively. (c) \didusig-grids of the Mn chain shown in panels (a) and (b) taken at energies indicated in the top right corner and by arrows in (b). White dashed lines mark the perimeter of the $\mathrm{Mn}_{18}~2a-[001]$ chain. White ticks perpendicular to the chain's main axis highlight the atomic positions in the chain, which are obtained from (a). Measurement parameters: 
    (a)-(c) $\vstab=\SI{-2.5}{\milli \volt}$, $\istab=\SI{1}{\nano \ampere}$ and $\vmod = \SI{20}{\micro \electronvolt}$.} 
\end{figure}
Motivated by the results of \cref{sec:001dimers}, where we demonstrated that a Mn dimer in \dooe~with an interatomic spacing of $2a$ shows hybridized and considerably split (order of \SI{100}{\micro \electronvolt}) YSR states and is, therefore, potentially interesting in terms of Majorana physics, we constructed such a chain consisting of 18 Mn atoms, which we refer to as $\mathrm{Mn}_{18}~2a-[001]$ chain in the following. An STM image of this chain is shown in \cref{fig:2a-001}(a). The chain is defect free and has an interatomic spacing of $2a$, while keeping a minimum distance of $\sim \SI{1}{\nano \meter}$ to oxygen impurities. A \didusig-line profile and the corresponding stabilization height profile are shown in \cref{fig:2a-001}(b). Besides the coherence peaks at $\pm \varDelta_\textrm{Ta}= \pm\SI{690}{\micro \electronvolt}$, we observe multiple pairs of states at lower energy, which are almost exclusively visible at negative biases. The highest energy one is located at $- \SI{500}{\micro \electronvolt}$. From a comparison of the stabilization height profile (blue curve) with the \didusig-line profile, we find that this state has intensity maxima on every atomic site of the $\mathrm{Mn}_{18}~2a-[001]$ chain, which are highlighted by two exemplary green dashed vertical lines. This is further supported by the \didusig-map in \cref{fig:2a-001}(c) taken at $- \SI{500}{\micro \electronvolt}$. Additionally, there is an increased \didusig-intensity at both ends and in the center of the chain, which could be interpreted as a standing wave with three maxima and might be a result of confined quasiparticles of a Shiba band \cite{Schneider2021b}.
\newline
In contrast, the state at $-\SI{400}{\micro \electronvolt}$ is located everywhere on the chain, as visible in the \didusig-line profile \cref{fig:2a-001}(b) and in the \didusig-map \cref{fig:2a-001}(c). As both states are located primarily inside the dashed ellipse which marks the perimeter of the Mn chain and as both have a strong particle-hole asymmetry (the high-intensity states are at negative biases), we conclude that both states have their origin in the single atom $\alpha$ state. It appears to form a comparably flat band, which extends from its band bottom at $-\SI{400}{\micro \electronvolt}$ to $\SI{-500}{\micro \electronvolt}$.
\newline
For positive biases we find a distinct spatial distribution at +\SI{400}{\micro \electronvolt}. We observe dumbbell-like states localized on every atomic site, which are oriented in the \deeo~with a nodal line at the longitudinal axis of the chain. As the spatial distribution matches the shape of YSR states of non-interacting $\beta$ states of a single atom (cf. \cref{fig:atomspec}(c)), we conclude that the $\beta$ state doesn't form a highly dispersive band.
\newline
A similar physical behavior is observed for the states located at lower energy, $\pm \SI{70}{\micro \electronvolt}$. We find a \didusig-intensity maximum located on every interatomic site of the chain, as indicated by yellow dashed vertical lines in \cref{fig:2a-001}(b). The spatial distribution of those states (\cref{fig:2a-001}(c)) again matches that of multiple non-interacting single Mn atoms positioned side-by-side. More precisely, the $\gamma$ state of a single Mn adatom resembles a $d_{xy}$-orbital, which in the case of an arranged array of adatoms in \dooe~with a $2a$-spacing adds up to orbital lobes oriented in the \deeo, located on every interatomic site. As this is what we observe in \cref{fig:2a-001}(c) (cf. intensity maxima and white ticks, which indicate atomic positions), we conclude that the $\gamma$ states don't form a dispersive band either.
\newline 
 Overall, in contrast to the expected physical properties of a $\mathrm{Mn}_{18}~2a-[001]$ chain as extrapolated from the behavior of the $2a-[001]$ dimer, we find non-interacting YSR states which do not couple strong enough to form dispersive bands in the case of $\beta$ and $\gamma$ states. If that was the case, we would expect to find standing waves of BdG quasiparticles confined by the geometry of the adatom chain (see e.g. \cite{Beck2022_2, Schneider2021b, Schneider2021a}). We observe a flat Shiba band for the $\alpha$-like states, which has a small width of $\sim \SI{100}{\micro \electronvolt}$. As this band is too narrow to cross the Fermi level, we conclude that $2a-[001]$ Mn chains are not interesting in terms of tailoring MBS. However, these properties indicate that next-nearest neighbor hoppings should play a minor role in the band formation of $1a-[001]$ Mn chains on Ta(110) (see Ref.\cite{Beck2022_2}), making them similar to the Mn/Nb(110) system \cite{Beck2021}.
\subsection{\label{sec:110dimers}Spectroscopic investigation of Mn dimers in \deeo}
\begin{figure*}
    \centering
    \includegraphics[scale=1]{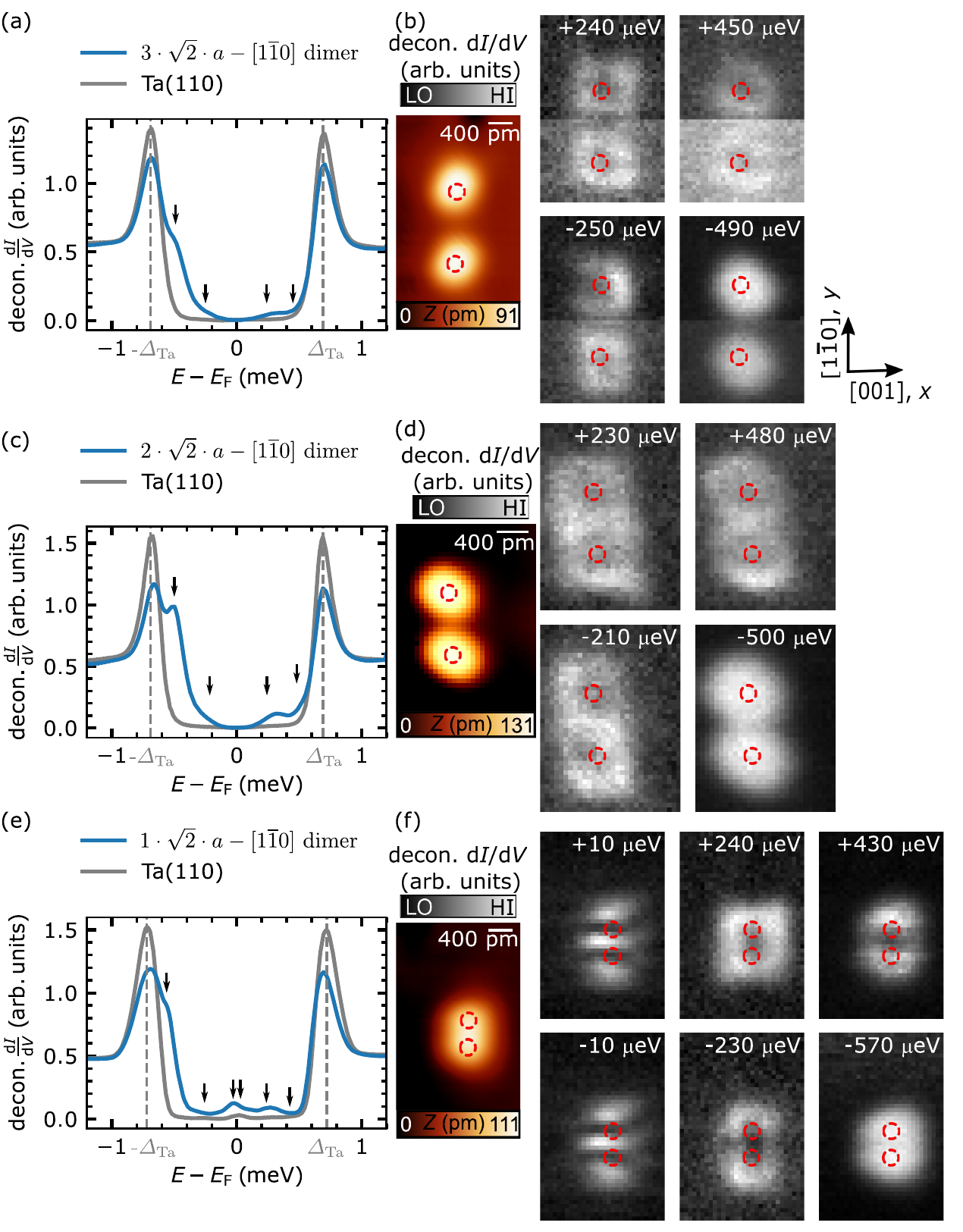}
    \caption{\label{fig:110dimers} Interatomic spacing-dependent investigation of artificial Mn dimers in \deeo~with separations of (a), (b) $3\sqrt{2}a$, (c), (d) $2\sqrt{2}a$ and (e), (f) $\sqrt{2}a$. Panels (a), (c) and (e) each show a deconvoluted \didu~of the Ta(110) substrate (gray) and a deconvoluted \didu~ averaged over the spatial extent of the respective dimer (blue), in order to capture all YSR states. Gray dashed vertical lines mark the superconducting gap of Ta(110). Panels (b), (d) and (f) display the spatial distributions of the YSR states, using spectroscopic grids (gray scale images). Black arrows in (a), (c) and (e) indicate the energetic positions, where the maps were measured. An STM image of the respective dimer is shown in the left of each panel (colored images). The crystallographic directions indicated in (b) are valid for all panels. Red circles highlight the positions of the Mn adatoms in the dimers. Measurement parameters: $\vstab=\SI{-2.5}{\milli \volt}$, $\istab=\SI{1}{\nano \ampere}$ and $\vmod= \SI{20}{\micro \volt}$.}
\end{figure*}
We proceed to discuss the physical properties of Mn dimers constructed in \deeo, with spacings of $3\sqrt{2}a$, $2\sqrt{2}a$ and $\sqrt{2}a$ in a similar fashion as for the \ooe-oriented dimers in \cref{sec:001dimers}. Spectroscopic results are shown in \cref{fig:110dimers} in the form of deconvoluted \didus~((a), (c) and (e)), which were averaged over the dimer area, and spatial distributions of YSR states ((b), (d) and (f)). Quite strikingly, we do not observe any splittings of YSR states for spacings $3\sqrt{2}a$ and $2\sqrt{2}a$. The \didus~of those two dimers, which are shown in \Cref{fig:110dimers}(a) and (c), match the \didu~of a single impurity quite accurately, c.f. \cref{fig:atomspec}(b). We find that there are minor shifts in the energies of the YSR states, compared to the single impurity. However, the overall amount of observed YSR states is unchanged. Further, the spatial maps of the two pairs of YSR states which we observe in \Cref{fig:110dimers}(b) and (d) match the shapes of a single Mn impurity (cf. \cref{fig:atomspec}(c)). Therefore, we conclude that the overlap of the YSR wavefunctions is not sufficient to hybridize and split the YSR states into symmetric and antisymmetric states for both interatomic spacings.
\newline
Moving on to the densely-packed dimer with a spacing of $\sqrt{2}a$ we find that the situation is changed. We observe that there are at least three pairs of YSR states inside the superconducting gap, which is particularly well visible for negative biases. A pair of states is observed at $\pm \SI{10}{\micro \electronvolt}$ and another in-gap state is found at $+\SI{430}{\micro \electronvolt}$. The former has a spatial distribution with an intensity maximum in the center of the dimer with two additional lobes oriented in the \deeo, while the latter is comprised of a nodal line in the center of the dimer and dumbbell-shape in the \deeo. A possible scenario would be that this pair of YSR states are symmetric and antisymmetric combinations of the single impurity $\beta$ state. At energies of approximately $\pm \SI{240}{\micro \electronvolt}$ we find a pair of YSR states, whose spatial distributions of the positive and negative bias state still resemble the single atom $\gamma$ state, but the one at negative energies has a nodal line inbetween the atoms. A similar behavior is found for the state at $-\SI{570}{\micro \electronvolt}$ which we assign to the $\alpha$ YSR state as it has a spatial distribution which matches the corresponding state of the single Mn impurity, c.f. \cref{fig:atomspec}(c). 
\newline
From this investigation we conclude that a formation of a wide-enough band to cross the Fermi level in chains in the \deeo~is only expected for an interatomic spacing of $\sqrt{2}a$ as the hybridization of the YSR states in dimers with larger interatomic spacing is negligible. An in-depth, length-dependent study of such chains is presented \cref{sec:110}. Spin-polarized STM measurements of chains in this direction are shown in \cref{sec:spstm110}.
\subsection{\label{sec:spstm110} \texorpdfstring{Magnetic structure of $\sqrt{2}a$-\eeo~Mn chains}{Magnetic structure of close-packed-[110] Mn chains}}
\begin{figure}
    \centering
    \includegraphics[scale=1]{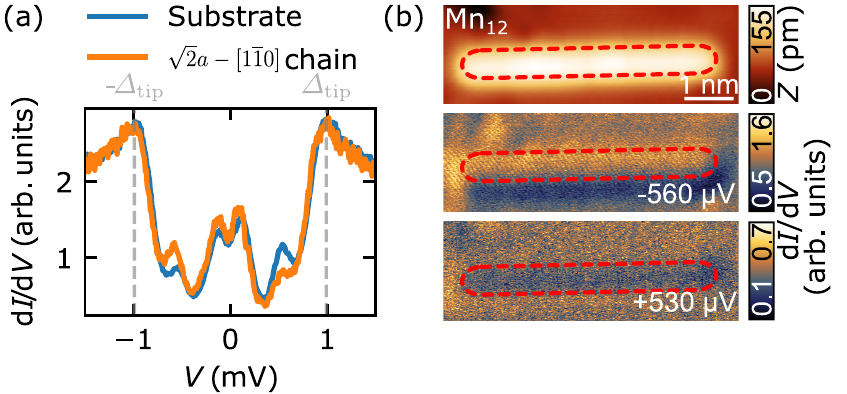}
    \caption{\label{fig:110_spstm}(a) Comparison of \didus~measured on the Ta(110) substrate and on a $\mathrm{Mn}_{12} \sqrt{2}a-[1\overline{1}0]$ chain using a superconducting Nb tip decorated with Mn atoms. The same microtip was used for all panels and an out-of-plane magnetic field of $+\SI{200}{\milli \tesla}$ was applied. (b) Topography and constant-contour \didusig-maps of a $\mathrm{Mn}_{12}$ $\sqrt{2}a-[1\overline{1}0]$ chain, measured at biases of the tips' YSR states. The dashed red lines mark the spatial extent of the Mn chain, as extracted form the topography. Measurement parameters: (a) $\vstab=\SI{1.5}{\milli \volt}$, $\istab=\SI{1}{\nano \ampere}$ and $\vmod = \SI{40}{\micro \volt}$, (b) $\vbias=\SI{5}{\milli \volt}$, $I=\SI{1}{\nano \ampere}$ and $\vmod = \SI{40}{\micro \volt}$.} 
    
\end{figure}

The magnetic ground state of Shiba chains has a crucial impact on their electronic properties and on their in-gap band structures. One can show that certain spin helices and their wave vector $\vec{k}_h$ transform to a Rashba-like spin-orbit coupling term \cite{Braunecker2010}, which again influences the induced $p$-type gap in a topologically superconducting phase \cite{Schneider2021b, Beck2022_2}, thereby reflecting the importance of knowing the magnetic ground state of YSR chains. In this section we gather information on the magnetic ground state of $\sqrt{2}a-[1\overline{1}0]$ Mn chains on Ta(110). 
\newline
To begin with, we perform SP-STM with YSR-functionalized superconducting Nb tips \cite{Schneider2021,Beck2022_2}. A \didu~measured on the bare substrate at an applied external magnetic field of $B=+\SI{200}{\milli \tesla}$, i.e. where the substrate's superconductivity is fully quenched, using a functionalized tip, is shown in \cref{fig:110_spstm}(a). We find two pairs of tip YSR states, one at $\pm \SI{110}{\micro \electronvolt}$ and one at $\pm \SI{530}{\micro \electronvolt}$. A spectrum measured on a $\mathrm{Mn}_{12}$ $\sqrt{2}a-[1\overline{1}0]$ chain displays an asymmetric intensity shift for the particle-hole partners of the YSR state at $\pm \SI{530}{\micro \electronvolt}$. As an effect of magnetoresistive tunneling \cite{Slonczewski1989, Wiesendanger2009}, the \didusig-signal decreases (increases) for the positive bias (negative bias) YSR state, comparing the \didu~measured on the substrate and on the Mn chain. Therefore, we conclude that the tip has a net spin-polarization at the biases of the pair of the tip's YSR states at $\pm \SI{530}{\micro \electronvolt}$. We verified this contrast by measurements over an AFM coupled $(\sqrt{3}/2) a-[1\overline{1}1]$ Mn chain, where alternating up-down contrast was easily achievable~\cite{Schneider2021}. 
\newline
Constant-contour \didusig-maps measured at those two biases are shown in \cref{fig:110_spstm}(b). We find that the \didusig-signal is evenly decreased (increased) for the positive bias (negative bias) state along the entire chain. This absence of any further contrast modulation indicates a constant out-of-plane magnetization of all Mn atoms in the chain.
\begin{figure}
    \centering
    \includegraphics[scale=1]{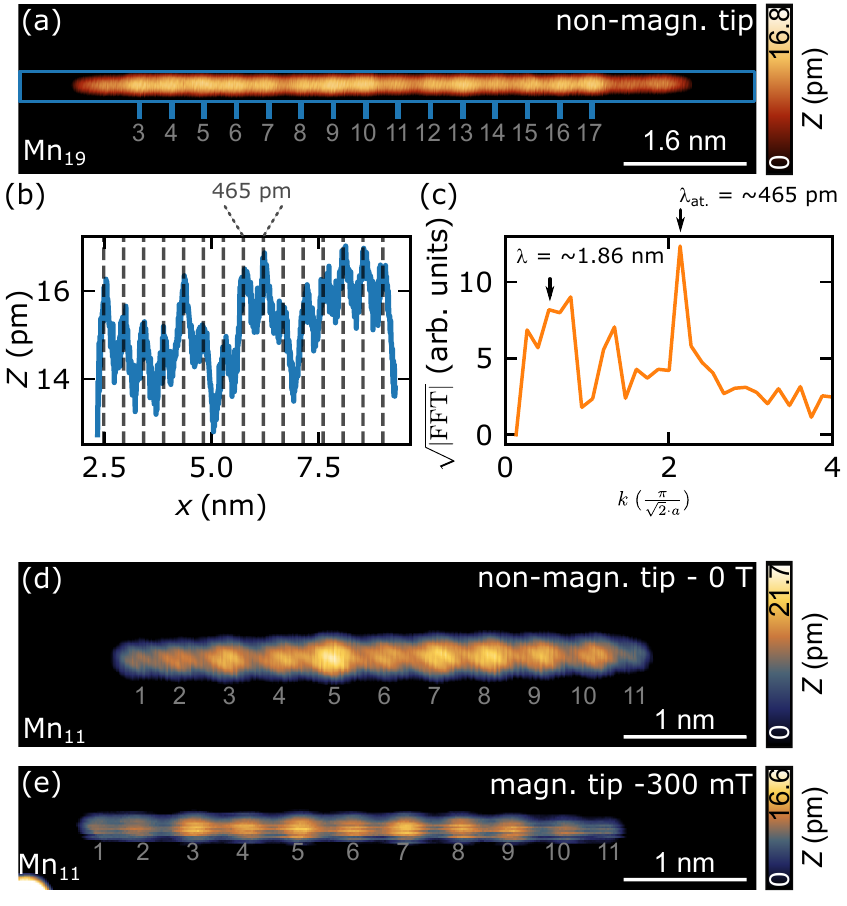}
    \caption{\label{fig:110_mod}(a) STM image of a $\mathrm{Mn}_{19}$ $\sqrt{2}a-[1\overline{1}0]$ chain taken with a non-magnetic tip. The contrast is adjusted to highlight the height-modulation of individual atoms in the chain. The blue rectangle indicates the extraction region for the line profile along the chain axis, which is shown in (b). Each $Z$-value of the line profile is obtained by averaging the pixels column wise, i.e. perpendicular to the chain axis. Grey dashed lines mark the positions of Mn atoms in the chain. (c) 1D FFT of the line profile shown in (b). Black arrows and labels mark the $k$-values of the interatomic distance in the chain and a broad peak around $k= 0.5 (\frac{\pi}{\sqrt{2}a})$. (d) and (e) are STM images of $\mathrm{Mn}_{11}$ $\sqrt{2}a-[1\overline{1}0]$ chains measured with a non-magnetic tip at $\SI{0}{\tesla}$ (d) and a magnetic tip at $\SI{-300}{\milli \tesla}$ (e), as indicated by the labels in the top right corner of each image, at different locations on the sample. Measurement parameters: (a) $\vbias=\SI{-20}{\milli \volt}$, $I=\SI{200}{\pico \ampere}$, (d) $\vbias=\SI{3}{\milli \volt}$, $I=\SI{1}{\nano \ampere}$, (e) $\vbias=\SI{2}{\milli \volt}$, $I=\SI{1}{\nano \ampere}$.}
\end{figure}

Additionally to these SP-STM measurements, we reproducibly observe a periodic modulation in the $Z$-signal on $\sqrt{2}a$-\eeo~Mn chains with different lengths, at multiple different locations on the sample and with different microtips and different bulk tips. In the following we discuss the obtained results. An example of this modulation is shown \cref{fig:110_mod}(a). The STM image was obtained with a regular superconducting tip (without YSR functionalization) and displays a $\mathrm{Mn}_{19}~\sqrt{2}a-[1\overline{1}0]$ chain. It is apparent, that the Mn atoms making up the chain have different heights. By bare eye, it is difficult to identify a certain wavelength of this modulation. However, one can find regions with a $4 \cdot \sqrt{2}a $ period.
We further analyze this by taking a line profile of the topography: The blue rectangle in \cref{fig:110_mod}(a) marks the extraction region where we averaged each line perpendicular to the chain axis to end up with the profile shown in \cref{fig:110_mod}(b). For the sake of visibility, we reduce the $x$-scale to the atoms 3-17, as the first two and the last two Mn atoms of the chain appear much lower than the rest of the chain.
\newline
In \cref{fig:110_mod}(b) one can clearly observe the atomic modulation with a nearest-neighbor distance of $\SI{465}{\pico \meter}$. In fact, the exact registry of the atomic distances in the chain with the underlying \deeo~of the Ta(110) substrate, is also observable in the STM image of \cref{fig:manipulation}(c) taken with an ultra-sharp tip. This precise and equal spacing of all nearest neighbors allows us to exclude an internal strain inside the chain as the origin of the height-modulation. Apart from the period of $\lambda_{at.} = \SI{465}{\pico \meter}$ it is more apparent from \cref{fig:110_mod}(b) that there is a modulation with $\lambda = 4 \cdot \SI{465}{\pico \meter}$. This additional modulation is observable in the 1D-FFT of the line profile, shown in \cref{fig:110_mod}(c), where one finds a broad peak around $k =\frac{1}{4} (\frac{2 \pi}{\sqrt{2}a})$.
\newline
To demonstrate that this modulation is reproducible, we used different microtips and different $\sqrt{2}a-[1\overline{1}0]$ $\mathrm{Mn}_{11}$ chains, constructed at different locations on the sample, and recorded the STM images shown in \Cref{fig:110_mod}(d) and (e). Further, we want to highlight that measurement (d) is taken in the superconducting state of the sample at $\SI{0}{\tesla}$, while image (e) is taken in the normal metal state of Ta(110) at $\SI{-300}{\milli \tesla}$. Comparing these two STM images, one finds that they show exactly the same contrast. Starting with the leftmost atom, we find that the first two and the last two atoms have a lower apparent height than atoms in the rest of the chain. For the 3$^\textrm{rd}$ atom from the left, the $z$-signal increases, until it reaches a maximum at site five. For atom six we observe a much lower apparent height, which increases greatly again for atom seven and then drops off slowly for sites eight and nine.
\newline 
Overall, we can not unambiguously determine the origin of this modulation. However, in the measurements shown above, we demonstrate that neither an out-of-plane magnetic structure, nor geometric anomalies or strain, nor electronic modulations caused by in-gap states (e.g. YSR states) can be the reason. Due to the independence of this contrast on the microtip and its existence with nonmagnetic tips, we speculate that weak in-plane non-collinear spin components in addition to the dominant ferromagnetic out-of-plane contribution (\cref{fig:110_spstm}) combined with tunneling anisotropic magnetoresistance could be it's origin \cite{Bode2002}. To confirm this hypothesis, one would have to conduct SP-STM with in-plane sensitive tips in a vector magnetic field or perform \textit{ab-initio} calculations of the magnetic ground state.
\subsection{\label{sec:110}\texorpdfstring{In-gap states of Mn chains in \deeo\ with $\sqrt{2}a$ spacing}{In-gap states of close-packed Mn chains in [110]-direction}}
\begin{figure}
    \centering
    \includegraphics[scale=1]{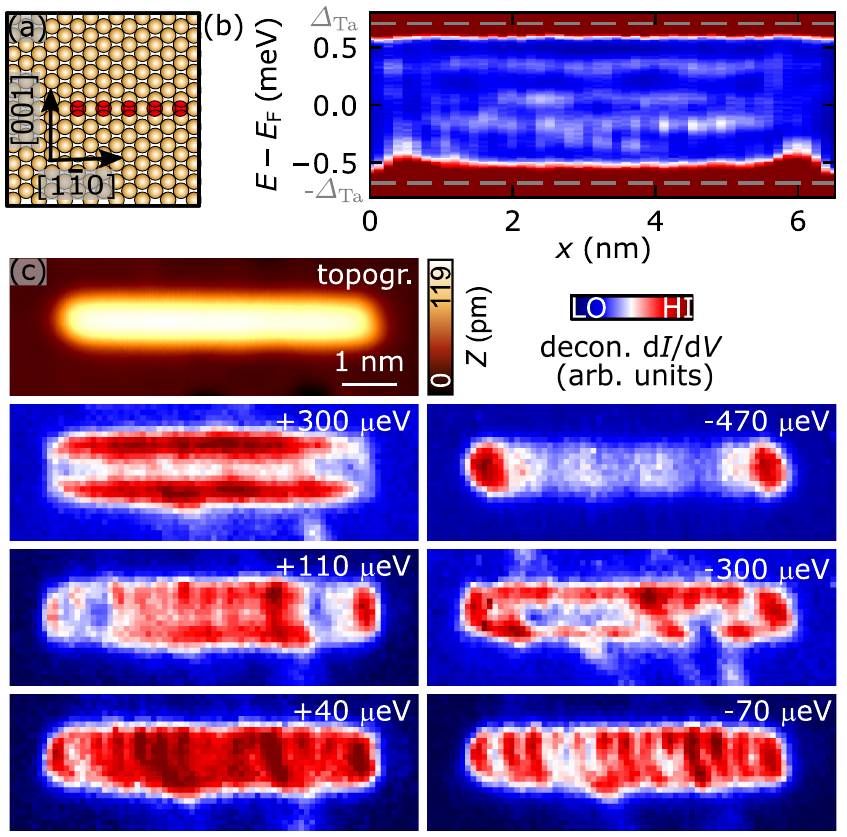}
    \caption{\label{fig:110}(a) Sketch of the arrangement of Ta atoms (yellow) and Mn atoms (red) for a $\sqrt{2}a-[1\overline{1}0]$ Mn chain. (b) \didusig-line profile measured along a $\mathrm{Mn}_{13}~ \sqrt{2}a-[1\overline{1}0]$ chain. Dashed gray horizontal lines mark the coherence peaks of $\varDelta_{\mathrm{Ta}}$. (c) STM image and \didusig-grid of a $\mathrm{Mn}_{13}$ chain evaluated at energy slices where confined states of the Shiba bands are observed. Measurement parameters: (b)-(c) $\vstab=\SI{-2.5}{\milli \volt}$, $\istab=\SI{1}{\nano \ampere}$ and $\vmod=\SI{20}{\micro \volt}$.}
\end{figure}
As identified by an in-depth study of dimers in \cref{sec:110dimers} and the spin-resolved STM measurements in \cref{sec:spstm110}, Mn chains on Ta(110) in \deeo~with an interatomic spacing of $\sqrt{2}a$ are potentially interesting concerning the formation of Shiba bands. An illustration of the building direction and the spectroscopic results of in-gap states are shown in \cref{fig:110}. An exemplary \didusig-line profile of a $\mathrm{Mn}_{13}$ chain is presented in \cref{fig:110}(b). We observe clearly defined in-gap states with a mirror symmetric spatial distribution around the center of the chain, indicating a good crystallographic quality and the absence of defects. However, while the states seem to have a standing wave character, the dispersion is not as readily observable as for $1a-[001]$ chains \cite{Beck2022_2}. 
\newline
\begin{figure}
    \centering
    \includegraphics[scale=1]{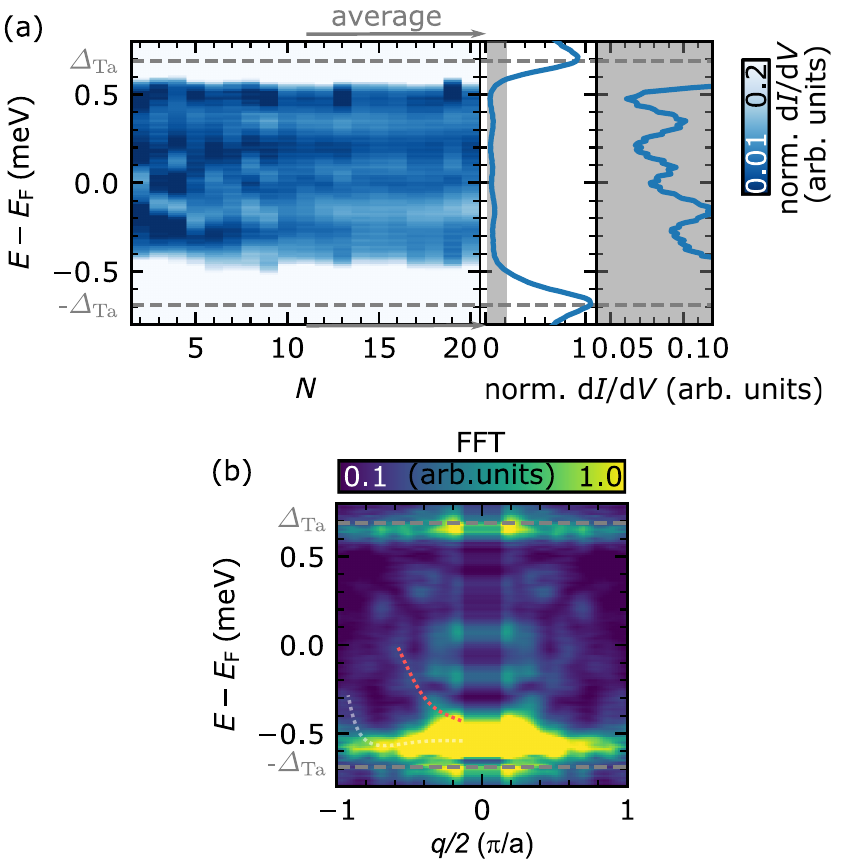}
    \caption{\label{fig:110_part2} (a) The leftmost panel displays chain length dependent \didus~for $\mathrm{Mn}_{2} -\mathrm{Mn}_{20}~\sqrt{2}a-[1\overline{1}0]$ chains. Each spectrum (column) was obtained by averaging over the \didusig-line profile similar to the one shown in \cref{fig:110}(b) of the particular length. An average \didu~ of the chains with lengths $\mathrm{Mn}_{11} -\mathrm{Mn}_{20}$ (indicated by the gray arrow) is shown in the middle panel. The rightmost panel is a magnification of the gray-shaded area in the middle panel to improve the visibility of states with a low \didusig~intensity. Gray dashed vertical lines mark the coherence peaks $\pm \varDelta_{\mathrm{Ta}}$. (b) Averaged 1D-FFT of \didusig-line profiles of Mn $\sqrt{2}a-[1\overline{1}0]$ chains with lengths $\mathrm{Mn}_{8}$-$\mathrm{Mn}_{20}$. The white and red dashed lines are guides to the eye. Measurement parameters: $\vstab=\SI{-2.5}{\milli \volt}$, $\istab=\SI{1}{\nano \ampere}$ and $\vmod=\SI{20}{\micro \volt}$.}
\end{figure}
The highest-intensity YSR state is found at $\SI{-470}{\micro \electronvolt}$ and is strongly localized on the edges of the chain (\cref{fig:110}(c)). We speculate that this trivial bound state could be the single impurity $\alpha$ state, which does not seem to hybridize and split (see \cref{fig:110dimers}(e) and (f)) in this chain geometry. Another state is observed at $\pm \SI{300}{\micro \electronvolt}$, which is located everywhere along the longitudinal axis of the chain and has the spatial distribution of a dumbbell-like YSR state in \dooe, as shown in the \didusig-maps of \cref{fig:110}(c). Further, we observe a state at $\SI{40}{\micro \electronvolt}$, which resembles a standing wave with four maxima along the chain, where the two inner ones are strongest. Additionally, we observe a state at $\SI{-70}{\micro \electronvolt}$, which is located on every atomic site. 
Overall, similar results are obtained for all chain lengths, as shown in \cref{fig:110_part2}(a) and Supplemental Movie 2. While details are difficult to extract, we can still conclude that there are dominant states at $-\SI{120}{\micro \electronvolt}$, $+\SI{100}{\micro \electronvolt}$ and $+\SI{300}{\micro \electronvolt}$ for chains with $N>8$. 
\newline
The averaged 1D-FFT of the \didusig-line profiles of all Mn chains with lengths of $\mathrm{Mn}_{8}$-$\mathrm{Mn}_{20}$ is shown in \cref{fig:110_part2}(b). We find indications for multiple bands: One has its band bottom at the $\Gamma$-point and $\SI{-550}{\micro \electronvolt}$. This Shiba band has an upwards dispersion, which is visible from its increase to $\sim \SI{-300}{\micro \electronvolt}$ close to the Brillouin zone boundary at $q/2 > 0.7~ (\frac{\pi}{a})$ (white dashed line). The band has a strong particle-hole asymmetry, as it is not observed for positive energies.
\newline
We find a second upwards dispersing Shiba band which has its band bottom at $q/2=0$ and $\SI{-400}{\micro \electronvolt}$. It has a more parabolically shaped dispersion than the first Shiba band and increases to $\SI{0}{\micro \electronvolt}$ at $q/2 \approx 0.5~ (\frac{\pi}{a})$, as indicated by the red dashed line. Therefore, we conclude that this Shiba band is gapless within our experimental resolution.
\newline
Overall, we find that chains in \deeo~with an interatomic spacing of $\sqrt{2}a$ have multiple bands with multiple different orbital origins (\cref{fig:110}(c) and \cref{fig:110_part2}(b)).
Unfortunately, a clear extraction of the structure of the Shiba bands including their respective orbital origins was not possible for $\sqrt{2}a-[1\overline{1}0]$ chains, due the limited energy resolution on the small-gap superconductor Ta ($T_\textrm{c}=\SI{4.39}{\kelvin}$) at our measurement temperature of $\SI{320}{\milli \kelvin}$ together with the multi-orbital character of the bands giving rise to a plethora of states in the gap.
\section{Conclusion}
In conclusion, we have demonstrated an improved method to prepare largely clean Ta(110), which enables not only the study of artificial 1D structures as in this work, but will also enable future studies of the self-organized growth of transition-metal films \cite{Menard2017, Palacio-Morales2019} on a superconductor, where the structures are exposed to high SOC. We have shown that STM tip-induced atom manipulation of Mn impurities on Ta(110) is very reproducible and precise. Using this method we established a measurement protocol to identify interesting building directions for 1D artificial structures and studied single Mn atoms and artificial dimers of multiple spacings in \dooe~and \deeo. 
\newline
We find that the $d_{\mathrm{z^2}}$-like YSR state of single Mn atoms on Ta(110) has very similar properties as on Nb(110). As this YSR state is crucial for the Shiba bands of $1a-[001]$ Mn chains on Ta(110) (\cref{sec:001dimers}) and on Nb(110) \cite{Schneider2021b} we have performed a comparative study on the influence of the substrate on Shiba band properties of $1a-[001]$ Mn chains in Ref.\cite{Beck2022_2}.
\newline
From the study of dimers in \deeo, we find that $\sqrt{2}a-[1\overline{1}0]$ chains could be promising candidates for 1D topological superconductors. We performed a length-dependent study of $\sqrt{2}a-[1\overline{1}0]$ Mn chains and investigated the Shiba band properties. However, we did not observe indications for a topological gap nor zero energy end states. As the YSR states which are spatially extended in the \deeo~clearly vary energetically for Mn adatoms on Ta(110) with respect to the Nb(110) substrate, $\sqrt{2}a-[1\overline{1}0]$ Mn chains on both substrates do not compare as easily as for the $1a-[001]$ chains. Additionally, we performed SP-STM measurements on $\sqrt{2}a-[1\overline{1}0]$ Mn chains, which indicate largely ferromagnetically aligned spins. However, we observe a contrast modulation of the apparent heights of the chain's atoms that might indicate additional non-collinear spin contributions in the surface plane.
\section*{Acknowledgments}
P.B., R.W., and J.W. gratefully acknowledge funding by the Deutsche Forschungsgemeinschaft (DFG, German Research Foundation) – SFB-925 – project 170620586. L.S., R.W., and J.W. gratefully acknowledge funding by the Cluster of Excellence 'Advanced Imaging of Matter' (EXC 2056 - project ID 390715994) of the DFG. R.W. gratefully acknowledges funding of the European Union via the ERC Advanced Grant ADMIRE (grant No. 786020).

%

\end{document}